\documentstyle[epsfig]{proceed}

\title[Interferometric observations of nearby galaxies]{Interferometric 
observations of nearby galaxies}
\author[Neininger] {N. Neininger\email{nneini@astro.uni-bonn.de}}
\institute{Radioastronomisches Institut der Universt\"at Bonn,
Auf dem H\"ugel 71, D-53121 Bonn, Germany\\
}

\begin{document}
\maketitle \abstract{This contribution presents a short overview about 
the investigation of the molecular gas content of nearby galaxies 
using the Plateau de Bure Interferometer.  I try to give a rather 
complete list of recent references, and explicit descriptions are 
given for objects under study at RAI Bonn.  They comprise very 
different types of galaxies such as Haro\,2, Mrk\,297, M\,82, 
NGC\,2146, M\,31 and NGC\,4258.  All galaxies had been studied with 
the IRAM 30-m telescope before and typically, the resolution into 
individual components by the PdBI revealed much more detail than 
expected.  This implies a note of caution for the interpretation of 
single-dish data in terms of velocity structure, opacity and virial 
mass determination.}


\section{Overview}

In the last few years the sensitivity and versatility of mm-wave 
interferometers have improved in large steps.  Using mosaicking 
techniques, it has become a relatively easy task to map the regions 
of the molecular gas emission in nearby galaxies with high angular 
and spectral resolution at a very good signal-to-noise ratio. 

Here, I will present some examples of results on nearby galaxies 
obtained recently that cover a relatively large range of galaxy types.  
All observations were performed at the IRAM Plateau de Bure 
interferometer (PdBI), the most sensitive telescope in 
the range of 80\ldots250 GHz.  The instrument is located in the French 
Alps at an altitude of 2550\,m. At present it consists  of 5 antennas 
of a diameter of 15\,m each which can be placed on different stations 
along a {\sf T}-shaped track.  Usually, the antennas are arranged in 
one of four standard configurations yielding baselines from 24\,m to 
230\,m (N-S) and 400\,m (E-W).  Typical angular resolutions start from 
about $4''$ for a compact array (called CD) at 100\,GHz and reach 
sub-arcsecond values for the most extended configuration.  For a more 
exhaustive description see Guilloteau et al.\ (1992) and  the IRAM 
website {\it http://iram.fr}\/.  
Due to the relatively large 
dishes the field of view is rather small, about $50''$ at 100\,GHz.  
This can however be overcome nowadays by applying mosaicking 
techniques and it remains the advantage of the larger collecting area.  
Properly set up, a mosaic covers the field of interest with an almost 
uniform sensitivity which is particularly useful for elongated sources 
such as outflows and edge-on galaxies.

The choice of the objects reflects the present work on external 
galaxies done in particular at the Radioastronomical Institute of the 
University of Bonn.  The focus is however a bit shifted towards 
particular objects that exploit the capabilities of the PdBI. Most of 
the projects are just started or in progress, so this is to a large 
extent a ``preview'' report.

\section{Dwarf and starburst galaxies}

The research on dwarf galaxies at the RAI Bonn is at the heart of a 
graduate school programme joining the Astronomical Institutes of Bonn 
and Bochum, supported by the Deutsche 
Forschungsgemeinschaft since 1993.  At its start aimed exclusively at 
the Magellanic Clouds, the topic has shifted towards low mass galaxies 
in general in a second period; the third period starting in 1999 also 
includes the comparison between dwarf and starburst galaxies. In this 
section four examples are shown, three of which are subject of a PhD
Thesis (T. Fritz, A. Wei\ss{}, A. Tarchi).

\subsection{A star forming dwarf}

{\em Investigators: T. Fritz, A. Heithausen, U. Klein, N. Neininger, 
C.L. Taylor, W. Walsh} \\[1ex]
Dwarf galaxies offer an excellent opportunity to probe the properties 
of the interstellar medium (ISM) in the absence of strong streaming 
motions, shear forces or large density gradients.  Despite of their 
low mass, some of them are actively star forming galaxies at a rate 
that is high compared to their gas mass.  These ``blue compact dwarf 
galaxies'' (BCDGs) are known to have a low metal abundance and almost 
no detectable molecular gas.  Several surveys have been conducted with 
the aim to detect CO in dwarf galaxies, but generally in vain.  Based 
on a H{\sc i} survey of Taylor et al.\ (1994), a more detailed search 
was carried out with the 30-m telescope (Barone et al., in preparation) 
and a clear detection of CO emission in the BCDG Haro\,2 could be 
obtained.  This was the basis of subsequent PdBI observations.

Haro\,2 was observed in the compact `CD' configuration as a mosaic of 
3 fields.  This was chosen in order to cover the entire star-forming 
body as it had been determined from optical photometry (Loose and 
Thuan 1986).  Extended emission of $^{12}$CO (1-0) and (2-1) was 
easily detected in the field, showing a relatively strong central 
peak and two lobes extended along the major axis of the star forming 
region.  This extent is consistent with the preceding 30-m 
observations, but now the structure can be investigated in much more 
detail.  Particularly surprising is the structure of the velocity 
field.  Haro\,2 is a dwarf galaxy with more or less elliptical 
appearance, but the emission of the molecular gas shows a very 
disturbed velocity field with steep gradients.  It is not even clear 
whether there are several separate components or maybe even a 
merger.  The nature of this unexpected behaviour is currently under 
investigation.

\subsection{A merging starburst ``dwarf''}
{\em Investigators: S. H\"uttemeister, U. Klein, N. Neininger, A. 
Greve} \\[1ex]
The upper size limit for dwarf galaxies is not well defined, but the 
Clumpy Irregular galaxy Mrk\,297 is certainly a bit at the bigger 
side with a mass of about $2\times10^{10}$ M$_{\odot}$. It also shows 
intense star formation. This one consists of two distinct kinematic 
components and is interpreted as an ongoing merger of two late-type 
spirals, one seen edge-on, the other face-on.  Mrk\,297 was covered 
with a four-field mosaic in the compact configuration and shows a 
rather concentrated emission of 
$^{12}$CO. The velocity structure is however rather complex over the 
range of detectable emission ($\sim 180$ km/s wide). We hope to get a 
closer view about the molecular gas distribution and its properties, 
in particular in the interaction zone.

\subsection*{Starburst galaxies}

If intense star formation occurs at the scale of a large part of a 
galaxy the phenomenon is called starburst galaxy.  At present, we are 
conducting studies of two prominent examples of this case, one being 
the ``prototype'' M\,82 and the other the special case of a starburst 
without obvious trigger, NGC\,2146.  In conjunction with the isolated 
star bursts in dwarf galaxies like those mentioned above or the 
Magellanic clouds, we hope to obtain a better understanding of the 
conditions of such an event.

\subsection{M\,82}
{\em Investigators: N. Neininger, A. Wei\ss{}, U. Klein, M. Gu\'elin, 
R. Wielebinski} \\[1ex]
The nearby irregular galaxy M\,82 is commonly called the prototype 
starburst galaxy.  Because of its brightness in the IR/mm regime and 
its proximity (3.25\,Mpc) a wealth of observations has been done to 
understand the origin and nature of the intense star formation 
activity.  The trigger of the activity seems to be clear: obviously, 
M\,82 is interacting with its neighbours M\,81 and NGC\,3077 and the 
tidal forces during the encounter are supposed to have started the 
exceptional star formation.  On the other hand, the evolution of the 
burst and many other parameters are not yet understood and even the 
structure of the disk is still a matter of debate.  The rotation 
curve, obtained from near-IR spectra and CO observations, indicates 
that M\,82 is indeed a disk galaxy.  Early single-dish CO observations 
(e.g.\ Loiseau et al.\ 1988) were interpreted to show a molecular ring 
similar to that of the Milky Way close to the centre, probably 
confining the active region and collimating the strong outflow 
(Shopbell 1998).

Those data with an a spatial resolution of at best 150\,pc were 
however inadequate to show any detail of the distribution of the 
molecular gas.  Interferometric observations of CO at somewhat better 
resolution (e.g.\ Lo et al.\ 1987) and others covering the lines of 
tracers of dense gas (Brouillet and Schilke 1993) indicate a patchy 
structure of the material.  The optical depth of CO being most 
probably high (Wild et al.\ 1992), we observed M\,82 in the presumably 
optically thin (1$\rightarrow$0) line of $^{13}$CO with the PdBI 
(Neininger et al.\ 1998b). 

In general, the spatial distribution of the $^{13}$CO is rather 
similar to that of the $^{12}$CO (Shen and Lo 1995): concentrated in 
two lobes embracing a weaker central region (Fig.~\ref{fig:m82-n2146}).  
In contrast to the 
low-resolution data which have commonly been interpreted as reflecting a 
torus of gas, the interferometer maps are better described by the 
presence of a molecular bar.  For the stellar population, such a bar 
had already been proposed by Achtermann and Lacy (1995).  A barred 
structure provides a straightforward means of transporting the gas 
needed to fuel the star formation towards the centre.  The weakness of 
the emission close to the nucleus (the active region) is certainly due 
to dissociation of the molecules in the strong radiation field.  In 
total, this provides a coherent global picture: during the passage 
close to M\,81, the tidal forces disrupted the outer disk of M\,82, 
leaving intergalactic streamers of atomic gas (Yun et al.\ 1994).  The 
remaining instable inner disk partially collapsed, thus starting the 
starburst acitivity that in turn provoked the strong wind.

The physical conditions in the actual disk remain however unclear.  
For example, the similarity of the distributions in the $^{12}$CO and 
$^{13}$CO line casts doubt on the determination of the optical depth.  
Moreover, the relationship between the individual energy sources and 
the conditions in the gas is not yet clear: A number of radio point 
sources has been identified (Kronberg et al.\ 1985) -- most of them 
are clearly identified as supernova remnants (SNRs) (see e.g.\ Wills 
et al.\ 1997 and references therein).  Around the strongest SNR we 
have identified a 130 pc-wide bubble which is characterized by warmer 
gas and an enhanced cosmic ray production rate.  It is however so big 
that this supernova could not possibly have created it -- the SNR just 
marks the position where previous SNe and stellar winds have created a 
particular environment. 

A number of questions remains, e.g.: what are the physical properties 
of the molecular gas in the central region of M\,82 -- such as the 
opacity, the temperatures, densities and abundances?  So we continued our 
studies by using the capability of the PdBI to perform dual-frequency 
observations; we observed the same field as before simultaneously in 
the (2-1) line of $^{12}$CO and in C$^{18}$O. The 
reduction and interpretation of these data is currently under way 
(Wei\ss{} et al., in preparation).

\subsection{NGC\,2146}
{\em Investigators: N. Neininger, A. Greve, A. Tarchi} \\[1ex] 
The ``dusty hand'' galaxy features a system of three dust lanes 
(spiral arms?)  and clear indications of a starburst such as a strong 
galactic wind.  But in contrast to other galaxies with strong star 
formation activity like M\,82 or NGC\,3628, no companion is visible 
that could have triggered the activity; and in contrast to Mrk\,297 
there is no obvious hint at a merger as well.  Nevertheless, the 
`hidden merger' scenario seems to be the most plausible explanation.  
The claim is that the starburst is indeed caused by a merging, but the 
encounter has happened long ago and no marked traces are left anymore.  
Looking for such traces thus is an additional task when studying the 
properties of the gas in the starburst region here.

\begin{figure}[htp]
\vbox{
\psfig{file=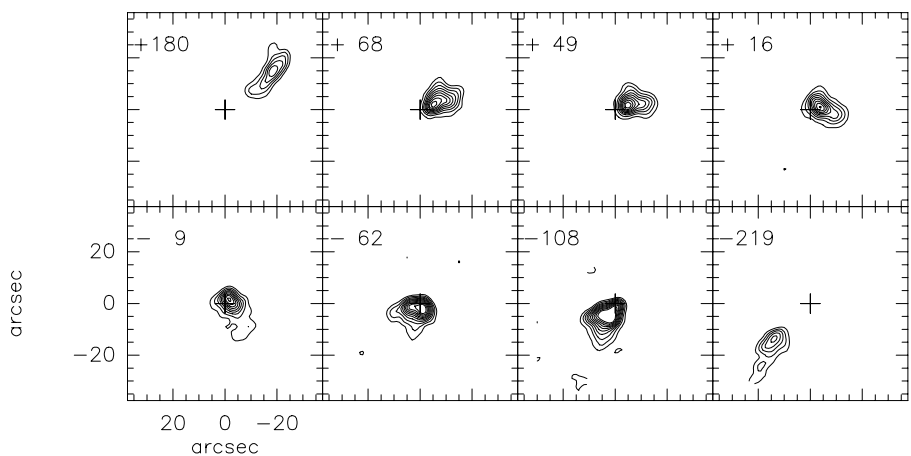,height=120mm,angle=270}\vspace{-51mm}}
\hfill
\parbox[b]{34.5mm}{
\caption{Some velo\-city channel maps of the $^{12}$CO(1-0) emission 
of NGC\,2146; the systemic 
velocity is subtracted. The cross marks the center of the galaxy. 
At all velocities except the extreme channels traces of an outflow
can be seen.}\label{fig:n2146chan}
}
\end{figure}

\begin{figure}[ht]
\vbox{
\epsfig{file=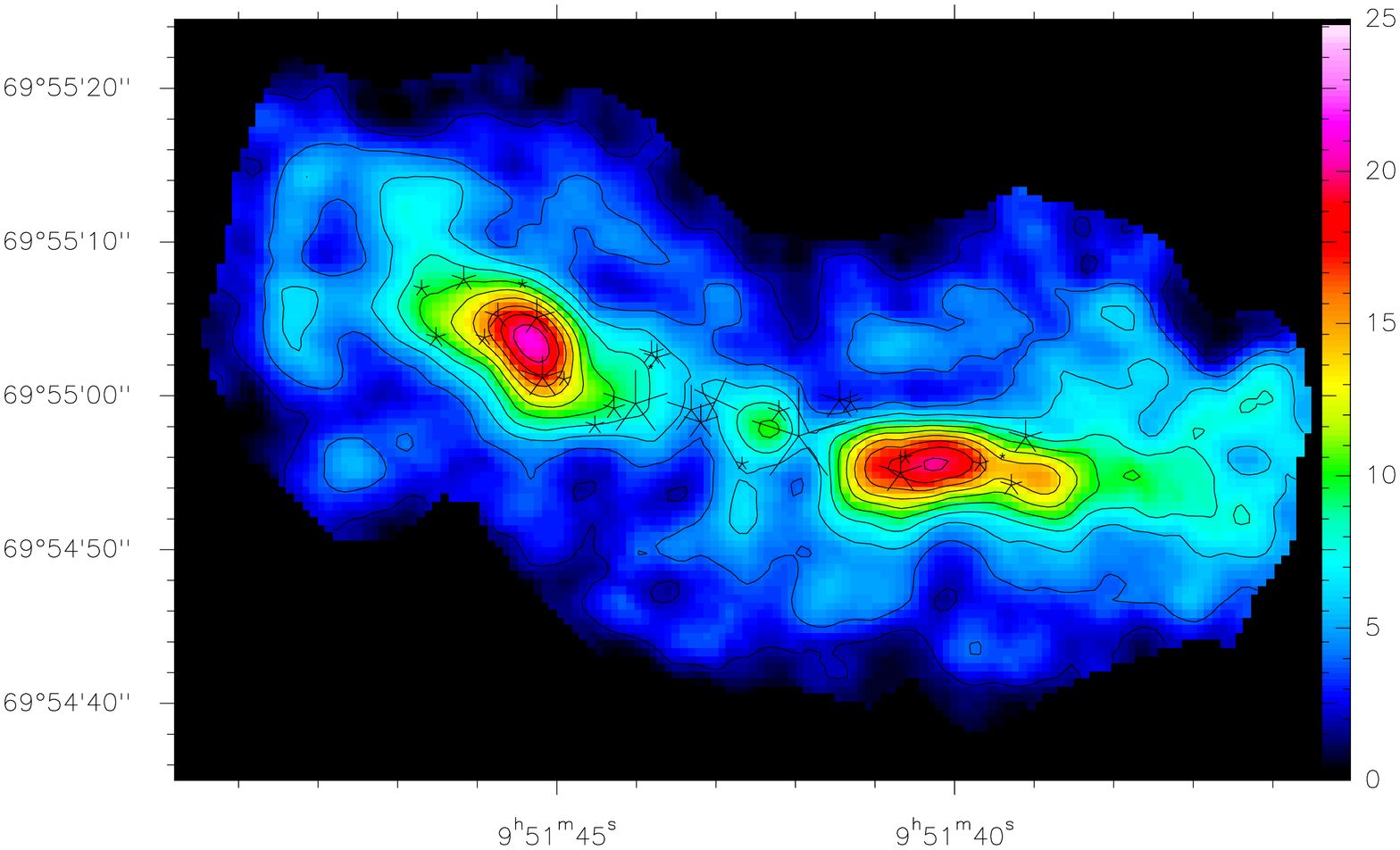,width=58mm,angle=270}\vspace{-63mm}
}\hfill
\epsfig{file=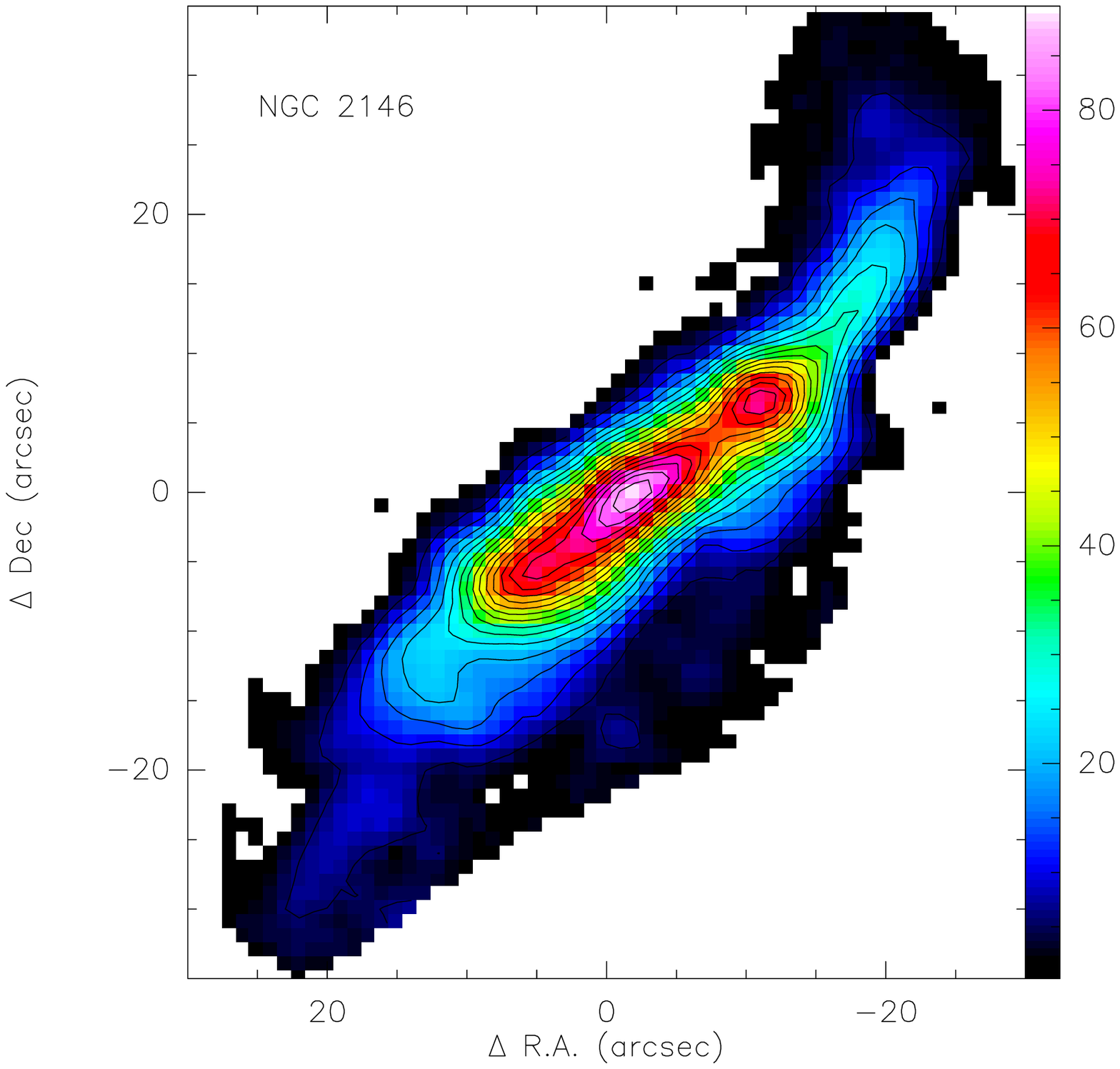,width=60mm,angle=270}\vspace{1ex}
\parbox{16cm}{
{\bf Fig. 3: Left:} The integrated intensity of the $^{13}$CO(1-0) emission 
of M\,82. The star symbols mark the positions of the radio point sources,
their size being proportional to the log of the radio flux. The main 
emission is situated in two lobes with different position angles whereas 
the central region is relatively weak. \\
{\bf Right:} The integrated intensity of the $^{12}$CO(1-0) emission of  
NGC\,2146. In contrast to the distribution in M\,82, it peaks towards the 
central region, where also the strongest radio point sources are found. 
Note also the warp already prominent at this small scale.}
\refstepcounter{figure}
\label{fig:m82-n2146}
\end{figure}

We mapped NGC\,2146 with the PdBI in the $^{12}$CO (1-0), (2-1) and 
the $^{13}$CO(1-0) emission lines, the parameters set in a way as to 
ensure a uniform coverage in all cases.  The detectable emission is 
well concentrated towards the center as in many other spiral galaxies.  
But already within these $60''$ (about 4 kpc) a warp is clearly 
visible (Fig.~\ref{fig:m82-n2146}, rightpart).  Consistent with the 
earlier findings 
no obvious hints at a second component are present, but the molecular
gas shows clear signs of an outflow (see Fig.~\ref{fig:n2146chan}).

Parallel to these observations of the molecular gas content, we have 
obtained high-resolution data at radio wavelengths ($\lambda$ 6 and 20 
cm) with a combination of MERLIN and the VLA. We could identify a 
number of point sources which are currently under investigation (Tarchi
et al., in prparation). The hope is to eventually compare their 
properties with their counterparts in M\,82.

\section{Normal galaxies}   

Especially since the availability of the mosaicking technique a 
growing number of nearby galaxies has been investigated. Results on 
several objects have been published recently and a (maybe incomplete) 
list of them is included in the references. Here I want to present 
the most ``extreme'' case in a little more detail.

\subsection{The highest spatial resolution: M\,31}
{\em Investigators: N. Neininger, M. Gu\'elin, R. Lucas et al.} 
\\[1ex]
\begin{figure}[ht]
\epsfig{file=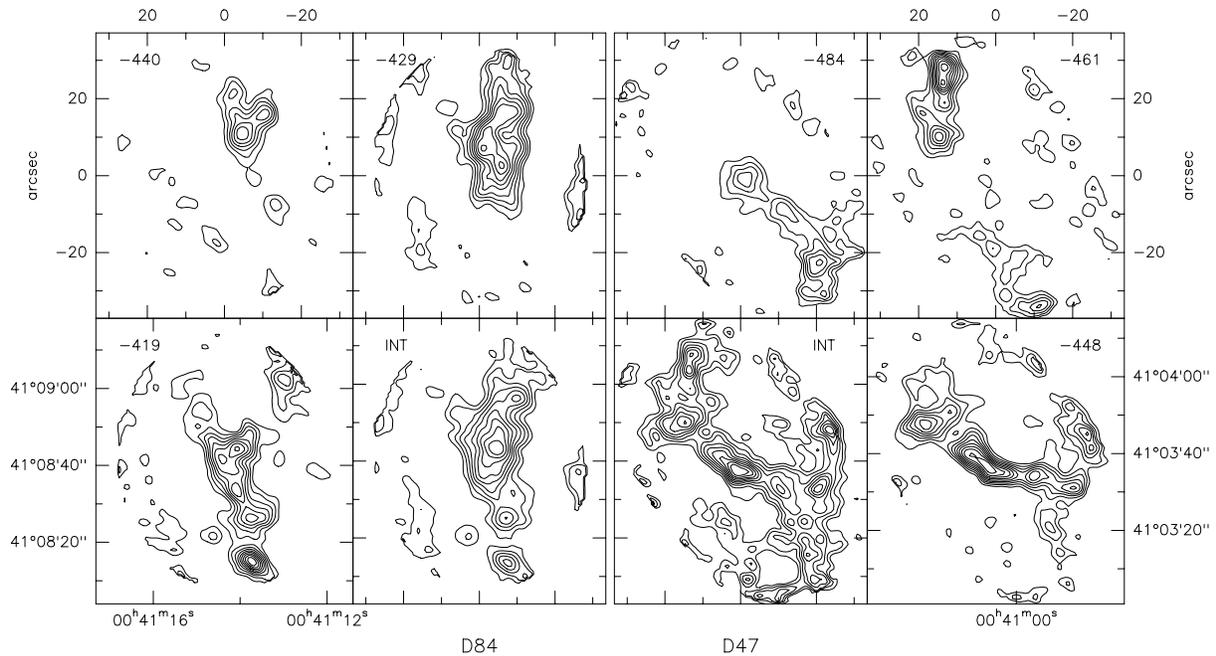,height=16cm,angle=270}
\caption{Comparative maps of two prominent cloud complexes in M\,31. 
The outer panels show velocity bins of a width of 10 km/s; the mean 
velocity is given in each field. The two lower panels marked `INT' give
the total intensity. In D84 the emission shows a continuous transition 
from one bin to the next; the corresponding spectrum is bright and 
narrow. D47 has a broad, double-peaked spectrum: e.g.\ in the center 
emission can only be found at the extreme velocities with a spacing of
about 40\,km/s, but not in the central channel.}
\label{fig:m31cloud}
\end{figure}
All the above described galaxies are several Mpc away from us and we 
can only derive global properties of the molecular gas content.  Even 
in our own Galaxy the detailed investigation of the molecular gas is 
limited to a few prominent examples and it is very difficult to obtain 
an homogeneous view at small and large scales.  A particularly tricky 
problem is the need to rely on velocity information to determine the 
distances of the objects under study and the related ambiguities.  
Therefore we started a global survey of the nearest grand design 
spiral, M\,31, with the 30-m telescope (Neininger et al.\ 1998a).  
From this survey we choose a number of cloud complexes for a further 
investigation with the PdBI. That way, we are able to combine a global 
view of the whole galaxy with detailed investigations at scale down to 
less than 10\,pc.

Among the chosen regions are obviously cold, quiescent clouds that 
show a bright, single-component emission line in the survey (e.g.\ 
situated in the dark cloud D84 -- the names are defined in Hodge 1981) 
as well as regions with multiple-peaked spectra.  One example of such 
a disturbed region lies in the dark cloud D47, about $5'$ or 1\,kpc 
away from D84 (see Fig.~\ref{fig:m31cloud} and Fig.\,2 in Neininger 
et al.\ 1998a).  The velocity separation in this complex is up to 
50\,km/s -- in the Galaxy, such a spectrum would be attributed to two 
components at very different places along the line of sight.  Here, it 
is clear from the location and the separation of the spiral arms that 
everything belongs to one cloud complex.  But wherefrom originate 
such big differences between those relatively close neighbours?  A 
comparison with the distribution of the H{\sc ii} regions gives a 
hint: the molecular cloud complex in D84 is isolated whereas that in 
D47 is located at the border of a particularly bright and extended 
H{\sc ii} region.  Similarly broad spectra are found in the big 
southern dark cloud D39 which hosts several star clusters.  This are 
only few examples, but they point all into the same direction: broad 
or multiple-component spectra are most likely caused by local effects.  
These cloud complexes are certainly not virialized on the scale of 
100\,pc, the resolution of the 30-m telescope at the distance of 
M\,31. The determination of the gas mass on the basis of data from the
two instruments yields grossly differing values. To further investigate 
the properties of the molecular cloud complexes in M\,31, we are 
enlarging our sample of combined studies with the two IRAM instruments 
while pushing the angular resolution well below the 10\,pc limit with 
the PdBI.

\section{A Black Hole candidate}
{\em Investigators: M. Krause, N. Neininger, C. Fendt} \\[1ex]
The spiral galaxy NGC\,4258 (or M\,106) started to become famous in 
the 1960's when it was recognized that its velocity field and the 
morphology of the H$\alpha$ emission was peculiar; further interest 
was rised by a high-resolution map of its radio emission (van 
der Kruit et al.\ 1972 and references therein). It shows two extended 
lobes that reach out along the minor 
axis of the optical image, with a steep (shock-?)  front and extended 
trailing plateaux -- if we assume for them a sense of rotation as 
defined by the main body of the galaxy.  Since then, different models 
for the origin of the anomalous features have been proposed.  The 
first one by van der Kruit et al.\ suggested an explosion in the 
center of the galaxy about 18 million years ago.  The galactic 
rotation then winds up the trail of the ejected material, thus forming 
the lobes. 

A new aspect was discovered in 1995 when Miyoshi et al.\ found 
evidence for a rapidly rotating accretion disk in the centre of 
NGC\,4258.  In the subsequent investigations it became one of the best 
candidates for a galaxy hosting a supermassive black hole.  This 
moreover implies the existence of a jet if we follow the actual 
picture of accretion disk systems.  Indeed, jets are known to create 
radio lobes and also the unusual H$\alpha$ arms might well be linked 
to a jet activity.  This scenario has a weak point, however: usually, 
it is assumed that jets are aligned with the rotation axis of the 
accretion disk which in turn is fixed in space by the black hole.  
The axis of rotation of the accretion disk is very close to the major 
axis of the galactic disk.  Thus, the jet has to travel a long 
way through the ISM of the galaxy.  Moreover, the galactic material is 
moving transversely through the path of the jet.  So it is necessary 
to investigate the properties of the molecular gas in NGC\,4258 in 
addition to the studies of the atomic hydrogen and the radio emission.

\begin{figure}[ht]
\vbox{
\epsfig{file=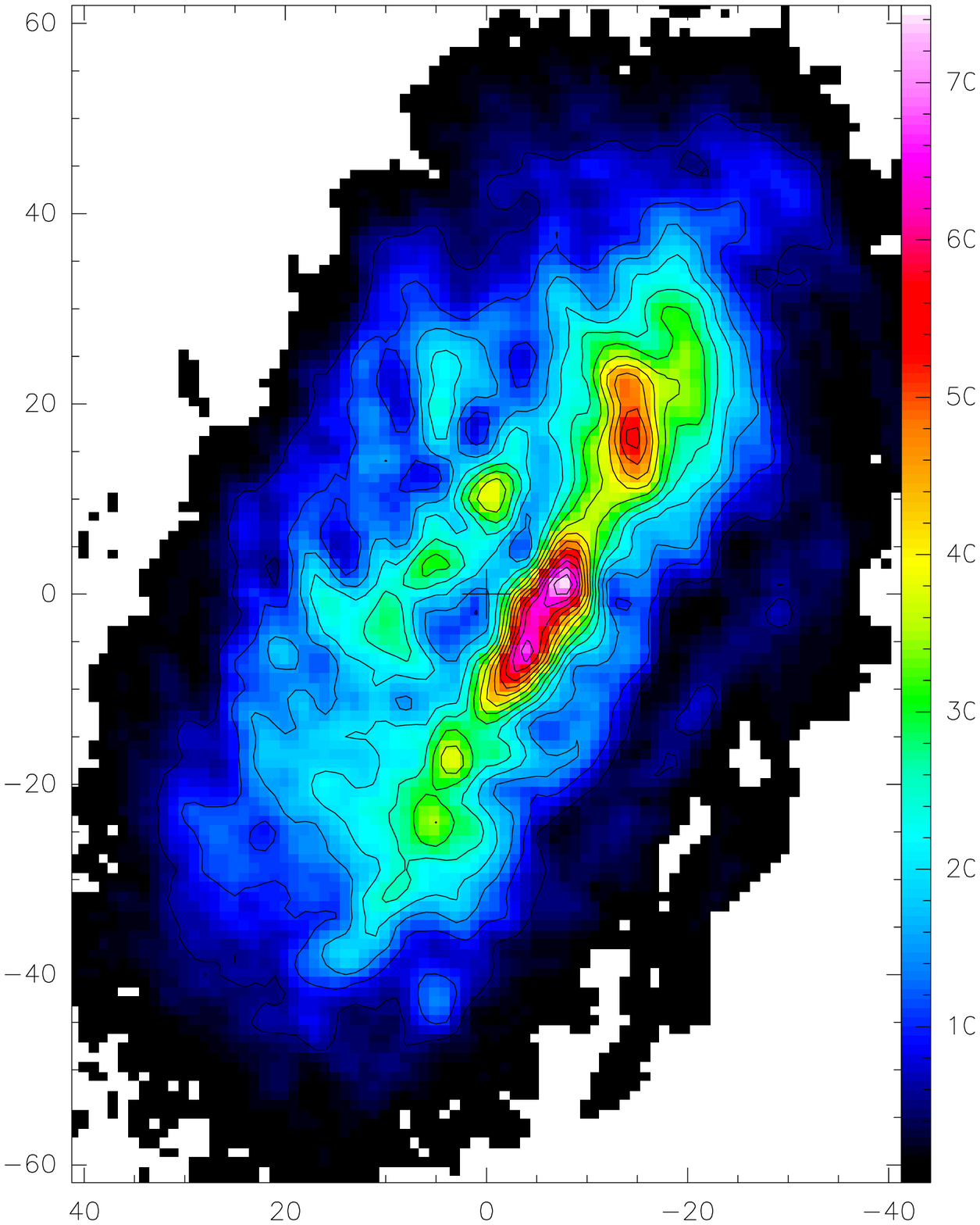,width=80mm,angle=0}\vspace{-45mm}}
\hfill
\parbox[b]{75mm}{
\caption{The $^{12}$CO(1-0) emission of NGC\,4258; the central accretion
disk is situated in the gap between the main emission ridge and the 
secondary ridge. The anomalous H$\alpha$ arms are situated in this gap
as well. The velocity structure of the molecular gas is very disturbed 
and far from normal rigid rotation. This is similar to the large-scale 
H{\sc i} velocity field (van Albada 1980), but not easily linked to the
H$\alpha$ velocity structure with its braided jet plus large-scale
rotation (Cecil et al.\ 1992).}
\label{fig:n4258int}
}
\end{figure}

Earlier 30-m observations had shown that the molecular gas is 
elongated along the anomalous H$\alpha$ arms (Krause et al.\ 1990) and 
hence we used a five-field mosaic to cover the whole emission region.  
The single emission ``bar'' (cf.\ Cox \& Downes 1996) seen by the 30-m 
telescope splits into two separate parts that form emission ridges on 
both sides of the anomalous diffuse H$\alpha$ arm 
(see Fig.~\ref{fig:n4258int}).  This suggests the 
existence of a tunnel with walls made of molecular gas which is filled
with hot ionized (atomic) gas that is entrained by the jet travelling 
along the axis of the tunnel. A similar scenario at a smaller scale is 
proposed for the outflows of Herbig-Haro objects (see e.g.\ Gueth et 
al.\ 1998 and references therein) where it can be more easily studied.

For NGC\,4258 the story is however not yet settled -- the evidence for 
the presence of a jet is accumulating, but the precise nature of it 
and the way how it may interact with the ISM is rather unclear. The 
PdBI data show a very distorted kinematical structure of the molecular 
gas (Krause et al.\ 1997) which is consistent with earlier H{\sc i} 
data (van Albada 1980) -- that way the local and the global kinematics 
seem to be linked.

\section{Summary}

Sensitive mm-wave interferometers have become a versatile tool to 
investigate even the relatively extended molecular gas emission 
of nearby galaxies. The gain in angular resolution may attain a 
factor of ten compared to single-dish instruments and quite often 
this really opens new views.  In particular the combination of 
high angular resolution and high sensitivity -- ideally combined 
with large-scale information from single-dish telescopes -- marks 
a major step forward. Most of the structures described or presented 
here were squeezed into a few spectra of the earlier observations
and hence difficult to interpret. The analysis of kinematic details 
or the structure of the molecular gas distribution and the
determination of its mass depends on such high-quality data.

\acknowledgements{It is a pleasure to thank my colleagues from the 
RAI for providing background information and excellent viewgraphs 
of their present work which I was allowed to present at the 
conference. }

\end{document}